%% file: main.tex
\errorstopmode
\input amssym.def
\input amssym.tex
\input format
\input epsf
\epsfclipon
\input macros
\input title

\input sect1

\input sect2

\input sect3

\input sect4

\input sect5

\input sect6

\input sect7

\input sect8

\input appa

\input appb

\input appc

\input biblio

\bye

%% file: format
% Page layout

\magnification=\magstephalf
\hsize=14.0 true cm
\vsize=19 true cm
\hoffset=1.0 true cm
\voffset=2.0 true cm

\abovedisplayskip=12pt plus 3pt minus 3pt
\belowdisplayskip=12pt plus 3pt minus 3pt
\parindent=1.0em

% Fonts

\font\sixrm=cmr6
\font\eightrm=cmr8
\font\ninerm=cmr9

\font\sixi=cmmi6
\font\eighti=cmmi8
\font\ninei=cmmi9

\font\sixsy=cmsy6
\font\eightsy=cmsy8
\font\ninesy=cmsy9

\font\sixbf=cmbx6
\font\eightbf=cmbx8
\font\ninebf=cmbx9

\font\eightit=cmti8
\font\nineit=cmti9

\font\eightsl=cmsl8
\font\ninesl=cmsl9

\font\sixss=cmss8 at 8 true pt
\font\sevenss=cmss9 at 9 true pt
\font\eightss=cmss8
\font\niness=cmss9
\font\tenss=cmss10

\font\sixmib=cmmib6
\font\sevenmib=cmmib7
\font\eightmib=cmmib8
\font\ninemib=cmmib9
\font\tenmib=cmmib10

 at 12 true pt
 at 12 true pt
\font\bigrm=cmr10 at 12 true pt
 at 12 true pt
 at 12 true pt

 at 16 true pt
%\font\Bigsy=cmsy12 at 16 true pt
%\font\Bigex=cmex12 at 16 true pt
 at 16 true pt
\font\Bigrm=cmr12 at 16 true pt
 at 16 true pt
 at 16 true pt

\catcode`@=11
\newfam\ssfam
\newfam\mibfam

\def\tenpoint{\def\rm{\fam0\tenrm}%
    \textfont0=\tenrm \scriptfont0=\sevenrm \scriptscriptfont0=\fiverm
    \textfont1=\teni  \scriptfont1=\seveni  \scriptscriptfont1=\fivei
    \textfont2=\tensy \scriptfont2=\sevensy \scriptscriptfont2=\fivesy
    \textfont3=\tenex \scriptfont3=\tenex   \scriptscriptfont3=\tenex
    \textfont\itfam=\tenit                  \def\it{\fam\itfam\tenit}%
    \textfont\slfam=\tensl                  \def\sl{\fam\slfam\tensl}%
    \textfont\bffam=\tenbf \scriptfont\bffam=\sevenbf
                           \scriptscriptfont\bffam=\fivebf
                           \def\bf{\fam\bffam\tenbf}%
    \textfont\ssfam=\tenss \scriptfont\ssfam=\sevenss
                           \scriptscriptfont\ssfam=\sevenss
                           \def\ss{\fam\ssfam\tenss}%
    \textfont\mibfam=\tenmib \scriptfont\mibfam=\sevenmib
                             \scriptscriptfont\mibfam=\sevenmib
                             \def\mib{\fam\mibfam\tenmib}%
    \normalbaselineskip=13pt
    \setbox\strutbox=\hbox{\vrule height8.5pt depth3.5pt width0pt}%
    \let\big=\tenbig
    \normalbaselines\rm}

\def\ninepoint{\def\rm{\fam0\ninerm}%
    \textfont0=\ninerm      \scriptfont0=\sixrm
                            \scriptscriptfont0=\fiverm
    \textfont1=\ninei       \scriptfont1=\sixi
                            \scriptscriptfont1=\fivei
    \textfont2=\ninesy      \scriptfont2=\sixsy
                            \scriptscriptfont2=\fivesy
    \textfont3=\tenex       \scriptfont3=\tenex
                            \scriptscriptfont3=\tenex
    \textfont\itfam=\nineit \def\it{\fam\itfam\nineit}%
    \textfont\slfam=\ninesl \def\sl{\fam\slfam\ninesl}%
    \textfont\bffam=\ninebf \scriptfont\bffam=\sixbf
                            \scriptscriptfont\bffam=\fivebf
                            \def\bf{\fam\bffam\ninebf}%
    \textfont\ssfam=\niness \scriptfont\ssfam=\sixss
                            \scriptscriptfont\ssfam=\sixss
                            \def\ss{\fam\ssfam\niness}%
    \textfont\mibfam=\ninemib \scriptfont\mibfam=\sixmib
                            \scriptscriptfont\mibfam=\sixmib
                            \def\mib{\fam\mibfam\ninemib}%
    \normalbaselineskip=12pt
    \setbox\strutbox=\hbox{\vrule height8.0pt depth3.0pt width0pt}%
    \let\big=\ninebig
    \normalbaselines\rm}

\def\eightpoint{\def\rm{\fam0\eightrm}%
    \textfont0=\eightrm      \scriptfont0=\sixrm
                             \scriptscriptfont0=\fiverm
    \textfont1=\eighti       \scriptfont1=\sixi
                             \scriptscriptfont1=\fivei
    \textfont2=\eightsy      \scriptfont2=\sixsy
                             \scriptscriptfont2=\fivesy
    \textfont3=\tenex        \scriptfont3=\tenex
                             \scriptscriptfont3=\tenex
    \textfont\itfam=\eightit \def\it{\fam\itfam\eightit}%
    \textfont\slfam=\eightsl \def\sl{\fam\slfam\eightsl}%
    \textfont\bffam=\eightbf \scriptfont\bffam=\sixbf
                             \scriptscriptfont\bffam=\fivebf
                             \def\bf{\fam\bffam\eightbf}%
    \textfont\ssfam=\eightss \scriptfont\ssfam=\sixss
                             \scriptscriptfont\ssfam=\sixss
                             \def\ss{\fam\ssfam\eightss}%
    \textfont\mibfam=\eightmib \scriptfont\mibfam=\sixmib
                             \scriptscriptfont\mibfam=\sixmib
                             \def\mib{\fam\mibfam\eightmib}%
    \normalbaselineskip=10pt
    \setbox\strutbox=\hbox{\vrule height7.0pt depth2.0pt width0pt}%
    \let\big=\eightbig
    \normalbaselines\rm}

\def\tenbig#1{{\hbox{$\left#1\vbox to8.5pt{}\right.\n@space$}}}
\def\ninebig#1{{\hbox{$\textfont0=\tenrm\textfont2=\tensy
                       \left#1\vbox to7.25pt{}\right.\n@space$}}}
\def\eightbig#1{{\hbox{$\textfont0=\ninerm\textfont2=\ninesy
                       \left#1\vbox to6.5pt{}\right.\n@space$}}}

\font\sectionfont=cmbx10
\font\subsectionfont=cmti10

\def\figurecaptionfont{\ninepoint}
\def\tablecaptionfont{\ninepoint}
\def\footnotefont{\eightpoint}

% New count registers

\newcount\equationno
\newcount\bibitemno
\newcount\figureno
\newcount\tableno

\equationno=0
\bibitemno=0
\figureno=0
\tableno=0
%\advance\pageno by -1

% Footline

\footline={\ifnum\pageno=0{\hfil}\else
{\hss\rm\the\pageno\hss}\fi}

% Section macro

\def\section #1. #2 \par
{\vskip0pt plus .10\vsize\penalty-100 \vskip0pt plus-.10\vsize
\vskip 1.6 true cm plus 0.2 true cm minus 0.2 true cm
\global\def\equationlabel{#1}
\global\equationno=0
\leftline{\sectionfont #1. #2}\par
\immediate\write\terminal{Section #1. #2}
\vskip 0.7 true cm plus 0.1 true cm minus 0.1 true cm
\noindent}

% Subsection macro

\def\subsection #1 \par
{\vskip0pt plus 0.8 true cm\penalty-50 \vskip0pt plus-0.8 true cm
\vskip2.5ex plus 0.1ex minus 0.1ex
\leftline{\subsectionfont #1}\par
\immediate\write\terminal{Subsection #1}
\vskip1.0ex plus 0.1ex minus 0.1ex
\noindent}

% Appendix macro

\def\appendix #1. #2 \par
{\vskip0pt plus .10\vsize\penalty-100 \vskip0pt plus-.10\vsize
\vskip 1.6 true cm plus 0.2 true cm minus 0.2 true cm
\global\def\equationlabel{\hbox{\rm#1}}
\global\equationno=0
\leftline{\sectionfont Appendix #1. #2}\par
\immediate\write\terminal{Appendix #1. #2}
\vskip 0.7 true cm plus 0.1 true cm minus 0.1 true cm
\noindent}

%\def\appendix #1. #2 \par
%{\vskip0pt plus .20\vsize\penalty-100 \vskip0pt plus-.20\vsize
%\vskip 1.6 true cm plus 0.2 true cm minus 0.2 true cm
%\global\def\equationlabel{\hbox{\rm#1}}
%\global\equationno=0
%\leftline{\sectionfont Appendix #1. #2}\par
%\immediate\write\terminal{Appendix #1. #2}
%\vskip 0.7 true cm plus 0.1 true cm minus 0.1 true cm
%\noindent}

% Displayed equations

\def\equation#1{$$\displaylines{\qquad #1}$$}
\def\enum{\global\advance\equationno by 1
\hfill\llap{{\rm(\equationlabel.\the\equationno)}}}
\def\noenum{\hfill}
\def\next#1{\cr\noalign{\vskip#1}\qquad}

% Bibliography macro, references

\def\ifundefined#1{\expandafter\ifx\csname#1\endcsname\relax}

\def\ref#1{\ifundefined{#1}?\immediate\write\terminal{unknown reference
on page \the\pageno}\else\csname#1\endcsname\fi}

\newwrite\terminal
\newwrite\bibitemlist

\def\bibitem#1#2\par{\global\advance\bibitemno by 1
\immediate\write\bibitemlist{\string\def
\expandafter\string\csname#1\endcsname
{\the\bibitemno}}
\item{[\the\bibitemno]}#2\par}

\def\beginbibliography{
\vskip0pt plus .15\vsize\penalty-100 \vskip0pt plus-.15\vsize
\vskip 1.2 true cm plus 0.2 true cm minus 0.2 true cm
\leftline{\sectionfont References}\par
\immediate\write\terminal{References}
\immediate\openout\bibitemlist=biblist
\frenchspacing\parindent=1.8em
\vskip 0.5 true cm plus 0.1 true cm minus 0.1 true cm}

\def\endbibliography{
\immediate\closeout\bibitemlist
\nonfrenchspacing\parindent=1.0em}

% Figure and table captions

\def\figurecaption#1{\global\advance\figureno by 1
\narrower\figurecaptionfont
Fig.~\the\figureno. #1}

\def\tablecaption#1{\global\advance\tableno by 1
\vbox to 0.25 true cm { }
\centerline{\tablecaptionfont%
Table~\the\tableno. #1}
\vskip-0.4 true cm}

\def\thicktablerule{\hrule height0.8pt}
\def\thintablerule{\hrule height0.4pt}

\tenpoint

\immediate\openin\bibitemlist=biblist
\ifeof\bibitemlist\immediate\closein\bibitemlist
\else\immediate\closein\bibitemlist
\input biblist \fi

% current year and month

\def\thismonth{\ifcase\month\or
January\or February\or March\or April\or May\or June\or
July\or August\or September\or October\or November\or December\fi}

%% file: macros
% Definitions and abbreviations

% Roman letters in math formulae

\def\rmd{{\rm d}}

\def\rme{{\rm e}}
\def\rmO{{\rm O}}

% Real and integer numbers

\def\Re{{\rm Re}\,}

% Special relations and symbols

\def\proof{\noindent{\sl Proof:}\kern0.6em}

\def\frac#1#2{\hbox{$#1\over#2$}}
\def\dual{\mathstrut^*\kern-0.1em}

\def\lvec#1{\setbox0=\hbox{$#1$}
    \setbox1=\hbox{$\scriptstyle\leftarrow$}
    #1\kern-\wd0\smash{
    \raise\ht0\hbox{$\raise1pt\hbox{$\scriptstyle\leftarrow$}$}}
    \kern-\wd1\kern\wd0}
\def\rvec#1{\setbox0=\hbox{$#1$}
    \setbox1=\hbox{$\scriptstyle\rightarrow$}
    #1\kern-\wd0\smash{
    \raise\ht0\hbox{$\raise1pt\hbox{$\scriptstyle\rightarrow$}$}}
    \kern-\wd1\kern\wd0}
\def\slash#1{\setbox0=\hbox{$#1$}\setbox1=\hbox{$\kern1pt/$}
    #1\kern-\wd0\kern1pt/\kern-\wd1\kern\wd0}

% Lattice derivatives

\def\nabstar#1{{\nabla\kern0.5pt\smash{\raise 4.5pt\hbox{$\ast$}}
               \kern-5.5pt_{#1}}}

\def\drvstar#1{{\partial\kern0.5pt\smash{\raise 4.5pt\hbox{$\ast$}}
               \kern-6.0pt_{#1}}}

\def\ldrvstar#1{{\lvec{\,\partial}\kern-0.5pt\smash{\raise 4.5pt\hbox{$\ast$}}
               \kern-5.0pt_{#1}}}

% Units

\def\MeV{{\rm MeV}}

\def\MSbar{\overline{\rm MS\kern-0.5pt}\kern0.5pt}

% Constants

% Fields

\def\psibar{\overline{\psi}}

\def\ubar{\bar{u}}
\def\dbar{\bar{d}}

% Dirac matrices

\def\dirac#1{\gamma_{#1}}
\def\diracstar#1#2{
    \setbox0=\hbox{$\gamma$}\setbox1=\hbox{$\gamma_{#1}$}
    \gamma_{#1}\kern-\wd1\kern\wd0
    \smash{\raise4.5pt\hbox{$\scriptstyle#2$}}}

% Gauge group

\def\Tr{{\rm Tr}}
\def\Ad{{\rm Ad}\kern0.1em}
\def\Str{{\rm Str}\kern1pt}

% Masses and decay constants

\def\mc{m_{\rm c}}
\def\mq{m_{\rm q}}
\def\mpi{M_{\pi}}
\def\Mpi{\mpi}

\def\Gpir{G_{\pi,\hbox{\sixrm R}}}

\def\mval{m_{\rm val}}
\def\mvalt{\tilde{m}_{\rm val}}
\def\Zg{Z_g}
\def\Zm{Z_m}
\def\Zmu{Z_{\mu}}
\def\ZA{Z_{A}}
\def\ZP{Z_{P}}
\def\ZS{Z_{S}}

\def\ZX{Z_{X}}

% Parameters and abbreviations

\def\Dm{D_m}
\def\Dmdag{{D_m}^{\kern-2pt\dagger}}

\def\csw{c_{\rm sw}}
\def\cA{c_{A}}
\def\bA{b_{A}}
\def\bP{b_{P}}
\def\bS{b_{S}}
\def\bPP{b_{P\kern-1pt P}}
\def\bPS{b_{P\kern-1pt S}}
\def\bg{b_g}
\def\bm{b_m}
\def\bmu{b_{\mu}}
\def\bX{b_{X}}
\def\bR{b_{R}}
\def\gr{g_{\hbox{\sixrm R}}}
\def\mr{m_{\hbox{\sixrm R}}}
\def\mur{\mu_{\hbox{\sixrm R}}}
\def\nur{\nu_{\hbox{\sixrm R}}}
\def\Mr{M_{\hbox{\sixrm R}}}
\def\Lr{\Lambda_{\hbox{\sixrm R}}}

\def\Seff{\Sigma_{\rm eff}}
\def\Mbar{{\kern1.5pt\overline{\kern-1.5ptM\kern-1pt}\kern1pt}}
\def\Sbar{\kern0.5pt{\overline{\kern-0.5pt\Sigma\kern-1pt}\kern1pt}_
          {\hbox{\sixrm R}}}
\def\Sval{\Sigma_{\rm val}}
\def\lbar{\bar{l}}
\def\mubar{\bar{\mu}}

% Rational approximation

\def\PM{{\Bbb P}_M}
\def\eps{\epsilon}
\def\seps{\sqrt{\epsilon}}
\def\Xop{{\Bbb X}}
\def\Mstar{M_{\ast}}
\def\nup{\nu'}
\def\omp{\omega_{+}}
\def\omm{\omega_{-}}
\def\ompm{\omega_{\pm}}
\def\delp{\Delta_{+}}
\def\delm{\Delta_{-}}
\def\delpm{\Delta_{\pm}}
\def\delt{\Delta_{0}}

% Chiral lagrangian

\def\Leff{{\cal L}}
\def\Mss{M_{\rm ss}}
\def\Mvv{M_{\rm vv}}
\def\Prop{G}

%% file: title
%
%\vbox{\vskip0.0cm}
\rightline{CERN-PH-TH/2008-239}

\vskip1.5cm 
\centerline{\Bigrm
Chiral symmetry breaking and the Banks--Casher relation}
\vskip0.3cm
\centerline{\Bigrm
in lattice QCD with Wilson quarks}
\vskip 0.6 true cm
\centerline{\bigrm Leonardo Giusti$\kern0.5pt\strut^{\rm a,b}$ and 
                   Martin L\"uscher$\kern0.5pt\strut^{\rm a}$}
\vskip1.5ex
\centerline{$\kern0.5pt\strut^{\rm a}$\kern0.0pt%
            {\it CERN, Physics Department, 1211 Geneva 23, Switzerland}}
\vskip0.5ex
\centerline{$\kern0.5pt\strut^{\rm b}$\kern0.0pt%
            {\it University of Milano-Bicocca and 
                 INFN Sezione di Milano-Bicocca, Milan, Italy}}
\vskip 0.8 true cm
\thintablerule
\vskip 2.0ex
\ninepoint
\leftline{\bf Abstract}
\vskip 1.0ex\noindent
The Banks--Casher relation links the spontaneous breaking of chiral
symmetry in QCD to the presence of a non-zero density
of quark modes at the low end of the spectrum of the Dirac operator.
Spectral observables like the number of modes
in a given energy interval are renormalizable and can therefore
be computed using the Wilson formulation of lattice QCD
even though the latter violates chiral symmetry at 
energies on the order of the inverse lattice spacing.
Using numerical simulations, we find (in 
two-flavour QCD) that the low quark modes 
do condense in the expected way. 
In particular, the chiral condensate
can be accurately calculated simply by counting 
the low modes on large lattices.
Other spectral observables can be considered as well
and have a potentially wide range of uses.
\vskip 2.0ex
\thintablerule

\tenpoint

\vskip-0.3cm

%% file: sect1
\section 1. Introduction

So far all results obtained in numerical lattice QCD are 
consistent with the expectation  
that chiral symmetry is spontaneously broken in the way presumed
by chiral perturbation theory. 
Little is known, however, about the dynamical processes
that cause the symmetry to break.
An intriguing remark, made long
ago by Banks and Casher [\ref{BanksCasher}], is that the effect
is tied to a condensation of the low modes of the Dirac
operator. Studies of the low modes may therefore provide important 
clues on the symmetry-breaking mechanism.

In the Wilson formulation of lattice QCD 
[\ref{Wilson}] and its improved versions [\ref{SW},\ref{OaImp}], 
chiral symmetry is violated
explicitly by terms proportional to the first or second power of the 
lattice spacing. The Banks--Casher relation consequently
cannot be expected to hold exactly
and the detailed properties of the low quark modes
could be significantly different from those in the continuum theory.
On the other hand, as long as only renormalizable quantities 
are considered, their values in the continuum limit must in principle 
be computable using the Wilson theory. 

The spectral density of the (hermitian) Dirac operator, 
and thus the average number of quark modes in a
given range of eigenvalues, are known to be renormalizable 
[\ref{Stability}]. 
In the present paper, we first give a second proof of this 
important fact (sect.~3). We then discuss the chiral 
perturbation expansion of the mode numbers and
show, in sect.~5, that their calculation
in lattice QCD requires only a modest computational effort.
Taken together, these results allow the chiral condensate
to be computed in the Wilson theory in a straightforward manner
(sect.~6). Spectral projectors however 
have a wider range of applicability and provide interesting
opportunities to explore the chiral regime of QCD,
some of which are briefly mentioned in sect.~7.

%% file: sect2
\section 2. Preliminaries

For simplicity we focus on QCD with a doublet of mass-degenerate
quarks, but the theoretical discussion 
is more generally valid and extends to the case of real-world QCD.
The quarks will be referred to as the up and down quarks, the associated
Goldstone bosons as the pions and the SU(2) flavour symmetry as the
isospin symmetry.
We consider both the continuum and the Wilson lattice theory 
in order to make it clear in which way
the mode number computed on the lattice is related to the one
defined in the continuum theory.

\subsection 2.1 Spectral density and mode number in the continuum theory

In a space-time box of volume $V$ with periodic or antiperiodic
boundary conditions, the euclidean massless Dirac operator $D$ in
presence of a given gauge field has purely imaginary eigenvalues
$i\lambda_1$, $i\lambda_2$, $\ldots$, which may be ordered so that
those with the lower magnitude come first. The associated 
average spectral density is given by
\equation{
  \rho(\lambda,m)={1\over V}\sum_{k=1}^{\infty}
  \left\langle\delta(\lambda-\lambda_k)\right\rangle
  \enum
}
where the bracket $\langle\ldots\rangle$ denotes the QCD expectation
value and $m$ the current-quark mass. Note that the isospin degeneracy
is not included in the mode counting, i.e.~the Dirac operator is
diagonalized in the subspace of, say, the up-quark fields.

The Banks--Casher relation [\ref{BanksCasher}]
\equation{
   \lim_{\lambda\to0}\lim_{m\to0}\lim_{V\to\infty}\rho(\lambda,m)
   ={\Sigma\over\pi}
   \enum
}
provides a link between the chiral condensate 
\equation{
  \Sigma=-\lim_{m\to0}\lim_{V\to\infty}\left\langle 
  \ubar u\right\rangle
  \enum
}
(where $u$ is the up-quark field) and the spectral density.
In particular, if chiral symmetry is spontaneously broken by
a non-zero value of the condensate, the density of the quark modes
in infinite volume does not vanish at the origin. 
A non-zero density conversely implies that the symmetry is broken,
i.e.~the Banks--Casher relation can be read in either direction.

Instead of the spectral density, the average number $\nu(M,m)$
of eigenmodes of the massive hermitian operator 
$D^{\dagger}D+m^2$ with eigenvalues $\alpha\leq M^2$
turns out to be a more convenient quantity to consider.
Evidently, since
\equation{
  \nu(M,m)=V\int_{-\Lambda}^{\Lambda}\rmd\lambda\,\rho(\lambda,m),
  \qquad
  \Lambda=\sqrt{M^2-m^2},
  \enum
}
the mode number ultimately 
carries the same information as the spectral density.

\subsection 2.2 O($a$)-improved lattice QCD

The lattice theory is set up as usual on
a hyper-cubic lattice with spacing $a$, time-like extent $T$ and
spatial size $L$. Periodic boundary conditions are imposed on
all fields and in all directions, the only exception being
the quark fields which are taken to be antiperiodic in time. 

As already mentioned, we focus on the Wilson theory 
in this paper. The details are not very relevant,
but for definiteness we choose the Wilson plaquette
action for the gauge field [\ref{Wilson}] and the standard expression
\equation{
  S_{\rm F}=a^4\sum_x\,\left\{\ubar(x)\Dm u(x)+\dbar(x)\Dm d(x)\right\}
  \enum
}
for the quark action, in which $\Dm$ denotes the massive, O($a$)-improved 
lattice Dirac operator [\ref{SW},\ref{OaImp}]. Apart from the bare
coupling $g_0$ and the bare mass $m_0$, the only free parameter 
in the lattice action is the improvement coefficient 
$\csw$, which we choose so as to cancel the O($a$)
lattice effects in on-shell quantities [\ref{NPimp}].

In this theory, 
the renormalized coupling and quark mass are related to the bare
parameters through [\ref{OaImp}]
\equation{
  \gr^2=\Zg(1+\bg a\mq)g_0^2,
  \enum
  \next{2.5ex}
  \mr=\Zm(1+\bm a\mq)\mq,\qquad
  \mq=m_0-\mc,
  \enum
}
where $\mc(g_0)$ denotes the critical bare mass and $\bg(g_0)$ and $\bm(g_0)$
are further O($a$)-improvement coefficients. 
The renormalization constants
$\Zg$ and $\Zm$
depend on the normalization conditions and 
are functions of the bare coupling and a normalization scale given in
units of the lattice spacing.

Composite fields like the isospin axial current and the isospin
pseudo-scalar and scalar densities are renormalized similarly by factors
of the form $\ZX(1+\bX a\mq)$ where $X=A,P,S$. 
The normalization conditions will be assumed to be such that the 
renormalized correlation functions satisfy the 
non-singlet chiral Ward identities up to terms of order $a^2$.
In particular,
\equation{
  \mr={\ZA(1+\bA a\mq)\over\ZP(1+\bP a\mq)}\,m+\rmO(a^2),
  \enum
}
where $m$ is the bare current-quark mass 
that appears in the PCAC relation [\ref{OaImp}].

On the lattice, we shall be mostly interested in 
the average number $\nu(M,\mq)$ of eigenmodes of $\Dmdag\Dm$ with eigenvalues
$\alpha\leq M^2$.
This definition of the mode number formally coincides with the one 
given in subsect.~2.1, but
it would evidently be premature to conclude 
that the values calculated on the 
lattice are simply related to the mode number defined in
the continuum theory.

%% file: sect3
\section 3. Renormalization of the mode number

The proof of the renormalizability of the mode number
given in this section
partly follows the lines of ref.~[\ref{Stability}],
but avoids some of the rather technical assumptions that had to be made there. 
An important new element of the proof is the use of
twisted-mass valence quarks and the associated density-chain 
correlation functions,
which have other applications as well (see sect.~7).

\subsection 3.1 Spectral sums and density chains

We consider the lattice theory and introduce the 
spectral sums
\equation{
  \sigma_k(\mu,\mq)=
  \bigl\langle\Tr\bigl\{\bigl(\Dmdag\Dm+\mu^2\bigr)^{-k}
  \bigr\}\bigr\rangle,
  \enum
}
where $k\geq3$ will be assumed for reasons to become clear below.
The spectral sums are related to 
the mode number $\nu(M,\mq)$ through the integral transform
\equation{
  \sigma_k(\mu,\mq)=
  \int_{0}^{\infty}\rmd M\,\nu(M,\mq)\,{2kM\over\left(M^2+\mu^2\right)^{k+1}},
  \enum
}
which can be shown to be invertible for every fixed $k$. 
The renormalization properties of $\nu(M,\mq)$ can therefore
be inferred from those of, say, $\sigma_3(\mu,\mq)$.

The inverse of the operator $\Dmdag\Dm+\mu^2$ coincides with the square
of the quark propagator in twisted-mass lattice 
QCD [\ref{tmQCD}].
We are thus led to add a set 
of isospin doublets 
$\psi_l$, $l=1,\ldots,2k$,
of valence-quark fields to the theory, with action
\equation{
  S_{\rm F,val}=a^4\sum_x\sum_{l=1}^{2k}\,
  \psibar_l(x)\left(\Dm+i\mu\dirac{5}\tau^3\right)\psi_l(x)
  \enum
}
(the isospin indices are suppressed in this formula and 
$\tau^3$ is the third isospin Pauli matrix).
Evidently, in order to cancel the valence-quark determinant,
a corresponding multiplet of pseudo-fermion fields 
must be added as well.
The spectral sums (3.1) can then be represented by density-chain 
observables like
\equation{
  \sigma_3(\mu,m)=-a^{24}\sum_{x_1,\ldots,x_6}
  \noenum
  \next{2.0ex}
  \hskip7.5em
  \bigl\langle P^{+}_{12}(x_1)P^{-}_{23}(x_2)P^{+}_{34}(x_3)P^{-}_{45}(x_4)
               P^{+}_{56}(x_5)P^{-}_{61}(x_6)\bigr\rangle,
  \enum
}
where $P^{\pm}_{ij}=\psibar_i\dirac{5}\tau^{\pm}\psi_j$ are the 
charged pseudo-scalar densities of the valence quarks (see
fig.~1).

\input figure1

\subsection 3.2 Renormalization of the spectral sums

With respect to the case of twisted-mass QCD discussed 
by Frezzotti et al.~[\ref{tmQCD},\ref{tmQCDimp}],
the O($a$)-improvement and renormalization of the partially quenched
theory considered here tends to be somewhat simpler.
In particular, we may choose a scheme
which is independent of the twisted mass parameter and
which coincides with the commonly used conventions
in the sea-quark sector of the theory.

At $\mu=0$, the Wilson theory
preserves the lattice symmetries, charge conjugation, the 
gauge symmetry and all (vector) flavour symmetries, including
the ones that mix the sea with the valence quarks.
Ultraviolet-divergent terms other than
those cancelled by the usual parameter and field renormalizations
are excluded by these symmetries.
When the twisted mass $\mu$ is switched on,
some of the symmetries are broken and further
ultraviolet-divergent terms can arise.
Power counting then shows that
a multiplicative renormalization,
\equation{
  \mur=\Zmu(1+\bmu a\mq)\mu,
  \enum
}
plus the renormalizations required at $\mu=0$ are
sufficient to renormalize the partially quenched theory.
Moreover, the correction proportional to $a\mq$ included in eq.~(3.5)
is all what needs to be added for on-shell O($a$)-improvement 
at $\mu\neq0$ [\ref{tmQCDimp}].

Considering eq.~(3.4), these remarks suggest that the
renormalization of 
$\sigma_3(\mu,\mq)$ is achieved by multiplication with the
sixth power of the renormalization factor $\ZP$ of the 
pseudo-scalar densities and by
renormalizing the parameters of the theory. The only worry one may
have at this point is that the summations in eq.~(3.4) over the
coordinates $x_1,\ldots,x_6$ diverge in the continuum
limit. However, as already pointed out in 
refs.~[\ref{GiustiRossiTesta},\ref{LuscherTop},\ref{Stability}], 
the short-distance singularities of density-chain correlation functions are
integrable, and give rise to
$\rmO(a\mq)$ corrections only, if there are six or more densities.

For any $k\geq3$, 
the renormalized O($a$)-improved spectral sums are thus
given by
\equation{
  \sigma_{k,\hbox{\sixrm R}}(\mur,\mr)=
  \left\{\ZP{1+\bP a\mq\over 1+\bPP a\mq}\right\}^{2k}\sigma_k(\mu,\mq),
  \enum
}
where it is understood that the bare masses 
are expressed through the renormalized ones.
The factors $1+\bPP a\mq$ in eq.~(3.6) are required for the 
cancellation of the $\rmO(a\mq)$ terms alluded to above 
which derive from the short-distance singularities
of the density-chain correlation functions [\ref{Stability}].

\subsection 3.3 Renormalized mode number

If the twisted-mass term is considered to be a perturbation 
of the theory at $\mu=0$, one quickly notices that
\equation{
  \Zmu=\ZP^{-1}
  \enum
}
is a possible (and natural) choice of the renormalization factor $\Zmu$.

Another simplification derives from the identity
\equation{
  {\partial\over\partial\mu}\sigma_k(\mu,\mq)=
  -2k\mu\sigma_{k+1}(\mu,\mq).
  \enum
}
When the renormalized spectral sums are similarly differentiated
with respect to the renormalized twisted mass $\mur$, the expressions
one obtains must be O($a$)-improved. As it turns out, this 
is the case if and only if
\equation{
  \bmu+\bP-\bPP=0.
  \enum
}
The renormalization factor in eq.~(3.6) thus becomes
\equation{
  \ZP{1+\bP a\mq\over 1+\bPP a\mq}=
  {1\over\Zmu(1+\bmu a\mq)}
  \enum
}  
up to terms of order $a^2\mq^2$.

Returning to the integral representation (3.2),
we now note that the renormalization factor $\{\Zmu(1+\bmu a\mq)\}^{-2k}$
needed to renormalize the spectral sum on the left of the equation
is cancelled on the right if we substitute
\equation{
  \Mr=\Zmu(1+\bmu a\mq)M
  \enum
}
and renormalize $\mu$.
We are thus led to conclude that
\equation{
  \nur(\Mr,\mr)=\nu(M,\mq)
  \enum
}
is a renormalized and O($a$)-improved quantity.
In other words, the mode number is a renormalization-group invariant.

\subsection 3.4 Universality

The steps taken in this section can be repeated using other
regularizations of QCD as long as these preserve 
the same (or more) symmetries
as the Wilson theory. Dimensional regularization with the 't Hooft--Veltman
prescription for $\dirac{5}$, for example, has all the required properties,
although in this case one is limited to weak-coupling 
perturbation theory.

Independently of the regularization, the renormalized mode number
will be the same if the same normalization conditions are used.
In particular, a definite convention such as the $\MSbar$ scheme
must be adopted
for the normalization of the pseudo-scalar densities.
The normalization of the sea-quark mass $\mr$ is then
determined by the PCAC relation, while the one of 
$\mur$ is fixed by requiring the identity
\equation{
  {\partial\over\partial\mur}\sigma_{k,\hbox{\sixrm R}}(\mur,\mr)=
  -2k\mur\sigma_{k+1,\hbox{\sixrm R}}(\mur,\mr)
  \enum
}  
to hold after removal of the regularization. At this point,
the renormalized spectral sums
are uniquely determined and so is the renormalized mode number, since
the integral transform
\equation{
  \sigma_{k,\hbox{\sixrm R}}(\mur,\mr)=
  \int_{0}^{\infty}\rmd\Mr\,
  \nur(\Mr,\mr)\,{2k\Mr\over\left(\Mr^2+\mur^2\right)^{k+1}}
  \enum
}
is free of normalization ambiguities.

%% file: figure1
\topinsert
\vbox{
\vskip0.0cm
\centerline{\epsfxsize=3.6cm\epsfbox{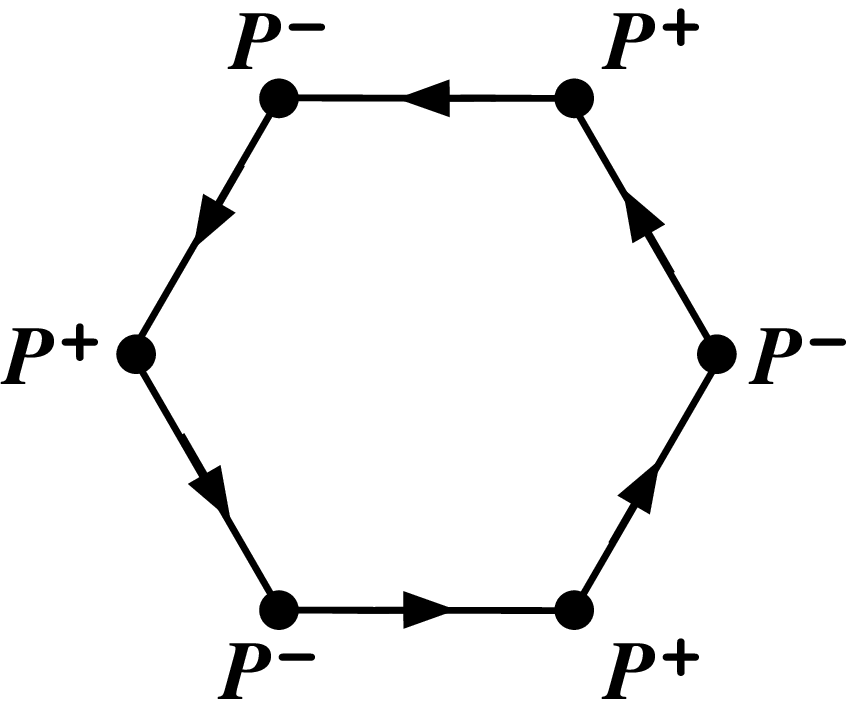}}
\vskip0.3cm
\figurecaption{%
The flavour labels of the pseudo-scalar densities in eq.~(3.4) are such that 
the contraction of the quark fields yields a 
closed quark loop with six edges. Each edge represents
a propagator $(\Dm\pm i\mu\dirac{5})^{-1}$ and each vertex 
contributes a factor $\dirac{5}$.
The ordered product of these factors,
summed over the positions $x_1,\ldots,x_6$ of the fields,
coincides with the trace (3.1).
}
%\vskip0.3cm
}
\endinsert

%% file: sect4
\section 4. Chiral expansion of the mode number

In the continuum theory and for small masses, the mode number
can be calculated analytically in chiral perturbation 
theory. Although all results quoted below are for the 
renormalized mode number, we omit the subscript ``R'' 
in this section in order to simplify the notation.

\subsection 4.1 Chiral perturbation theory

At present the chiral expansion of the 
spectral density $\rho(\lambda,m)$ is known
to next-to-leading order of chiral perturbation theory.
The first computation to this order was 
performed by Smilga and Stern [\ref{SmilgaStern}]
in the massless theory in infinite volume.
Later Osborn et al.~[\ref{OsbornEtAl}] and 
Damgaard et al.~[\ref{DamgaardEtAl}]
performed a more complete and
systematic computation based on partially quenched chiral perturbation
theory [\ref{BernardGolterman},\ref{SharpeShoresh}].

The starting point 
in the paper of Osborn et al.~is the formula
\equation{
  \rho(\lambda,m)={1\over2\pi}\lim_{\epsilon\to0}
  \left\{\Sigma_{\rm val}(\epsilon+i\lambda)+
         \Sigma_{\rm val}(\epsilon-i\lambda)\right\},
  \enum
}
which relates the spectral density to 
the expectation value $-\Sigma_{\rm val}(m_{\rm val})$
of the scalar density of an added valence quark of mass 
$m_{\rm val}$. With a doublet of sea quarks,
the relevant graded flavour symmetry group
is then ${\rm SU}(3|1)$ and the chiral expansion of 
$\Sigma_{\rm val}(m_{\rm val})$ is derived
from the associated chiral effective theory
(see appendix A).

\subsection 4.2 Large-volume regime

In infinite volume, chiral perturbation theory yields an expansion of 
$\rho(\lambda,m)$ essentially in powers of $\lambda$ and $m$.
The leading-order term is given by the Banks--Casher formula
and the ``effective chiral condensate'', defined through
\equation{
  \Seff={\pi\over2}{\nu(M,m)\over\Lambda V},
  \enum
}
therefore coincides with $\Sigma$ in the chiral limit.

At next-to-leading order, the chiral expansion reads
\equation{
  \left.{\Seff\over\Sigma}\right|_{V=\infty}=
  1-{m\Sigma\over16\pi^2F^4}\biggl\{
  3\ln{\Lambda\Sigma\over\mubar^2F^2}-3\lbar_6-1+
  \ln2+\ln\Bigl(1+{m^2\over\Lambda^2}\Bigr)
  \noenum
  \next{2ex}
  {\phantom{
  \left.{\Seff\over\Sigma}\right|_{V=\infty}=
  1-{m\Sigma\over16\pi^2F^4}\biggl\{
  }}
  +{m\over\Lambda}\arctan{\Lambda\over m}
  +{\Lambda\over m}\arctan{m\over\Lambda}\biggr\}+\ldots
  \enum
}
The constants $F$ and $\lbar_6$ in this expression are, respectively,
the pion decay constant in the chiral limit and
an ${\rm SU}(3|1)$ 
low-energy effective coupling renormalized at scale $\mubar$ (appendix A). 
Following the tradition [\ref{GasserLeutwyler}], $\mubar$ may be set
to the physical charged-pion mass, but since only the
scale-invariant sum of the first two terms in the curly bracket
matters, this choice is not compulsory.

\input figure2

A remarkable feature of eq.~(4.3) is that
the one-loop correction 
vanishes, for any value of $\Lambda$, when the quark mass goes to zero.
Smilga and Stern [\ref{SmilgaStern}] already noted the absence
of terms proportional to $\Lambda$ 
and showed that this
was a special property of the 
two-flavour theory.
The chiral corrections to $\Seff/\Sigma$ consequently tend to be 
quite small (see fig.~2 for illustration).

\subsection 4.3 Finite-volume effects

In the present context, the kinematical situation of interest
is the so-called $p$-regime of QCD,
where $T\geq L$, $FL\geq1$ and $m\Sigma V\gg F^2L^2$.
Chiral perturbation theory is easily extended 
to this regime and can be used to estimate the effects 
of the finite volume
[\ref{GasserLeutwylerFV}].

In the case of $\Seff$, the calculation shows that the dependence on the 
volume sets in at one-loop order and that 
the infinite-volume limit is reached at an exponential rate 
according to
\equation{
   \Seff-\left.\Seff\right|_{V=\infty}\propto\rme^{-{1\over2}M_{\Lambda}L},
   \qquad
   M_{\Lambda}^2={2\Lambda\Sigma\over F^2}.
   \enum
}
Note that $M_{\Lambda}$ coincides with the leading-order expression
for the mass of a pseudo-scalar meson made of two valence quarks
of mass $\Lambda$. 
Since $\Lambda$ is normally taken to be significantly 
larger than the sea-quark mass, the finite-size 
effects (4.4) tend to be smaller than 
those expected for the pion mass $\Mpi$, for example, which 
decrease like $\rme^{-\Mpi L}$. In particular, if the 
parameter values previously used in fig.~2
are inserted, and if $L\geq2$ fm is assumed,
$\Seff$ is estimated to deviate from its infinite-volume value
by a fraction of percent at most.

%% file: figure2
\topinsert
\vbox{
\vskip0.0cm
\centerline{\epsfxsize=8.0cm\epsfbox{plots/Seff.eps}}
\vskip0.3cm
\figurecaption{%
Quark-mass dependence of $\Seff/\Sigma$ at 
fixed $\Lambda$ according to next-to-leading order
of chiral perturbation theory.
The low-energy constants have been set to $\Sigma=(250\,\MeV)^3$,
$F=90\,\MeV$, $\mubar=140\,\MeV$ and $\lbar_6=3$ in this plot.
}
\vskip0.3cm
}
\endinsert

%% file: sect5
\section 5. Counting the low modes in lattice QCD

In presence of a given gauge field, the number 
of eigenmodes of $\Dmdag\Dm$ with eigenvalues $\alpha\leq
M^2$ can be determined straightforwardly by calculating
the eigenvalues and their multiplicities numerically.
The effort required for such computations
however grows proportionally to the second or perhaps even
a higher power of the space-time volume $V$. In this section, we
show that the modes can be counted more efficiently using
spectral projectors.

\subsection 5.1 Stochastic representation of the mode number

Let $\PM$ be the orthogonal projector to the subspace of quark fields
spanned by the eigenmodes of $\Dmdag\Dm$ with eigenvalues $\alpha\leq M^2$.
An alternative representation of the mode number
\equation{
  \nu(M,\mq)=\langle\Tr\{\PM\}\rangle
  \enum
}
is then given by
\equation{
  \nu(M,\mq)=\langle{\cal O}_N\rangle,
  \qquad
  {\cal O}_N={1\over N}\sum_{k=1}^N\left(\eta_k,\PM\eta_k\right),
  \enum
}
where we have added a set 
of pseudo-fermion fields, $\eta_1,\ldots,\eta_N$, 
to the theory with action
\equation{
  S_{\eta}=\sum_{k=1}^N\left(\eta_k,\eta_k\right).
  \enum
}
In the course of a numerical simulation, these fields
are generated randomly, for each gauge-field configuration,
and the mode number is estimated in the usual way 
by averaging the observable ${\cal O}_N$ over the generated
ensemble of fields.

The variance of ${\cal O}_N$,
\equation{
  \bigl\langle({\cal O}_N-\langle{\cal O}_N\rangle)^2\bigr\rangle
  =\bigl\langle(\Tr\{\PM\}-\langle\Tr\{\PM\}\rangle)^2\bigr\rangle
  +{1\over N}\nu(M,\mq),
  \enum
}
is larger than the one of $\Tr\{\PM\}$, but the difference can
be reduced
by increasing the number $N$ of pseudo-fermion fields.
More important may be the fact that the mode number is 
an extensive quantity, while
the variance of $\Tr\{\PM\}$
does not appear to grow with the volume $V$ of the lattice
at the values of $M$ of interest
[\ref{LuscherPalombi}]. 
At fixed $N$ and for a given statistics, the relative statistical error 
of the calculated mode number is therefore expected to 
decrease like $V^{-1/2}$.

\subsection 5.2 Rational approximation

The projector $\PM$ can be approximated fairly easily by rational
functions of $\Dmdag\Dm$. There are different ways to proceed and the
choices made in the following may not be the best ones, but the
proposed method is quite efficient and numerically safe.

Let $P(y)$ be the minmax polynomial of degree $n$ which
minimizes the deviation
\equation{
  \delta=\max_{\eps\leq y\leq 1}\left|1-\sqrt{y}P(y)\right|.
  \enum
}
The numerical 
computation of this polynomial for specified values of $n$
and $\eps>0$
is a standard task in 
approximation theory 
(see ref.~[\ref{NumMethods}], for example).
In the range $-1\leq x\leq 1$, the function
\equation{
  h(x)=\frac{1}{2}\left\{1-xP(x^2)\right\}
  \enum
}
then provides an approximation to the step function $\theta(-x)$.
By construction, the approximation error is at most
$\frac{1}{2}\delta$ if $|x|\geq\seps$ and
numerical inspection moreover shows
that $h(x)$ decreases monotonically in the transition 
region $|x|\leq\seps$.

An approximation to the projector $\PM$ is now given by
\equation{
  \PM\simeq h(\Xop)^4,
  \qquad
  \Xop=
  1-{2\Mstar^2\over\Dmdag\Dm+\Mstar^2},
  \enum
}
where $\Mstar\simeq M$ is an adjustable mass parameter.
The quality of the approximation is determined
by the values of $n$, $\eps$ and the ratio $M/\Mstar$.
In practice,
the degree $n$ of the minmax polynomial should be reasonably
small and the deviation
\equation{
  \Delta=\langle\Tr\{\PM-h(\Xop)^4\}\rangle
  \enum
}
must be much smaller than the
statistical errors of the calculated mode numbers.

The estimation of $\Delta$
and the choice of $M/\Mstar$ are discussed
in appendix B.
Here we only note that the computation of  
\equation{
  (\eta,\PM\eta)\simeq(\eta,h(\Xop)^4\eta)=\|h(\Xop)^2\eta\|^2
  \enum
}
requires the application of the square of $h(\Xop)$ to the 
pseudo-fermion field $\eta$ and not of its fourth power.

\subsection 5.3 Numerical implementation

The minmax polynomial $P(y)$ and therefore the operator $h(\Xop)$ 
can be expanded in a series of Chebyshev polynomials with 
rapidly decreasing coefficients [\ref{NumMethods}].
Chebyshev series of this kind can be safely 
evaluated using the Clenshaw recursion [\ref{Recipes}].

The computation of $h(\Xop)\eta$ for a given source
field $\eta$ then requires the operator $\Xop$ to be applied
$2n+1$ times. Each application essentially amounts
to solving the linear system
\equation{
  (\Dmdag\Dm+\Mstar^2)\psi=\eta
  \enum
}
using one's favourite iterative algorithm. 
This system is normally significantly better 
conditioned than the lattice Dirac equation
$\Dm\psi=\eta$.
Moreover, it is our experience that
a fairly loose stopping criterion can be chosen without 
compromising the correctness of the simulation results.

We finally remark that the computational effort 
required for the calculation of the mode number along the 
lines explained here scales like $V$ or at most $V\ln(V)$
as the lattice is increased.

%% file: sect6
\section 6. Computation of the chiral condensate

The simulations discussed in this section have a limited scope,
but the results clearly show
that the low modes of the Dirac operator condense and
that the mode number can be accurately computed
using the stochastic method described in the previous section.

We have considered two lattices in these studies, 
with spacing $a\simeq0.08$ fm, spatial 
sizes $L\simeq1.9$ fm and $2.5$ fm, respectively, and
time-like extents $T=2L$. 
The exact parameter values and 
further technical details are given in appendix C.
All values quoted for the renormalized mass parameters,
the mass-dependent condensate $\Sbar$ defined in subsect.~6.3
and the condensate $\Sigma$ refer to the $\MSbar$ scheme at $2$ GeV.

\input figure3

\subsection 6.1 Qualitative behaviour of the mode number

The data plotted in fig.~3 show that 
the mode number is, in the case considered, a nearly linear 
function of $\Mr$ from above the threshold region at 
$\Mr\simeq\mr$ up to at least $110$ MeV. 
This behaviour is qualitatively in line with 
chiral perturbation theory,
but the fact that the linear regime extends to such 
large values of $\Mr$
is rather striking and could not be foreseen.

At the very low end of the spectrum, the curve shown in fig.~3 however
clearly deviates from its expected form in the continuum theory
(shaded area in fig.~3) [\ref{Stability}]. 
A plausible explanation of the observed 
deviation is that chiral symmetry is not
exactly preserved in the Wilson theory and that the fine structure of
the spectrum of the Dirac operator near the threshold at $\Mr=\mr$ is
consequently not protected from perturbing lattice effects
[\ref{Sharpe}].
The deviation must in any case be a lattice artefact, since
the renormalized mode number is bound to converge to 
its continuum value as the lattice spacing is decreased
(cf.~sect.~3).

In the following, we focus on the linear regime in fig.~3, 
where the mode number is not expected
to be particularly sensitive to discretisation errors.
Moreover, since the effort required for the numerical 
calculation of the low eigenvalues of $\Dmdag\Dm$
is not small, the mode number was normally computed
using the method described in sect.~5 and we 
shall, from now on, only
discuss results obtained in this way.

\subsection 6.2 Volume-dependence of the mode number

In the large-volume regime of the theory, $\nu(M,\mq)/V$
is expected to be independent of the lattice size
up to exponentially small corrections (cf.~sect.~4).
The lattices we have simulated are such that
we can immediately check whether these
corrections are significant at the level of the statistical
errors.

To this end, we form the ratios 
\equation{
   r_{3,4}={\nu(M,\mq)_{{\rm D}_3}\over \nu(M,\mq)_{{\rm E}_4}}
   \left({32\over24}\right)^4,
   \qquad
   r_{5,5}={\nu(M,\mq)_{{\rm D}_5}\over \nu(M,\mq)_{{\rm E}_5}}
   \left({32\over24}\right)^4,
   \enum
}
where the subscripts $D_3$ etc.~refer to the run label 
quoted in table~2 (appendix C). 
Both ratios turn out to be practically equal to $1$.
More precisely, $r_{3,4}$ differs from $1$ by $-0.6$ to $-2.0$
standard deviations and $r_{5,5}$ by $+0.7$ to $+1.5$ standard
deviations as $M$ varies over the values listed in table~2.
There are thus no indications for
significant finite-volume effects on these lattices.

\subsection 6.3 Calculation of $\Sigma$

The values of the renormalized mode number which we calculated
on the larger of the two lattices considered are plotted
in fig.~4 (left graph).
At fixed quark mass, the mode number is, to a very good approximation, 
a linear function of $\Mr$ in the range shown 
in the figure. In particular,
the slope of the data can easily be determined
by quadratic interpolation (lines in the left graph).

\input figure4

We are thus led to introduce the mass-dependent condensate
\equation{
  \Sbar={\pi\over2V}\sqrt{1-\left({\mr\over\Mr}\right)^2}
  {\partial\over\partial\Mr}\nur(\Mr,\mr),
  \enum
}
where the prefactor is chosen such that $\Sbar$ coincides with
the chiral condensate $\Sigma$ to leading order of 
chiral perturbation theory. In table~1 we list the 
calculated values of $\Sbar$ at $\Mr=95$ MeV
(a point in the middle of the available range of masses).
The first errors quoted in the table are the statistical ones,
while the second errors are those inherited
from the product of the lattice spacing and 
the renormalization factors needed to convert from lattice
to physical normalizations (appendix C).

\input table1

The next-to-leading order formula (4.3) suggests 
that $\Sbar=\Sigma$ up to higher-order corrections and
terms vanishing proportionally to $\mr$
in the chiral limit.
Note that there are no terms 
proportional to $\mr\ln\mr$ at this order of the chiral expansion. 
The data for $\Sbar$ at $\Mr=95$ MeV actually fall on a straight
line (right graph in fig.~4)
and the extrapolation to $\mr=0$ then yields the estimate 
\equation{
  \Sigma^{1/3}=276(3)(4)(5)\,\MeV
  \enum
}
for the chiral condensate. Higher-order 
corrections were neglected here, but appear to be small as the 
results vary only little (within roughly the third error 
given above) when 
the chiral limit is taken at other values of $\Mr$.

It goes without saying, however, that this procedure
and the quoted result for the condensate 
will have to be confirmed by more extensive 
calculations. Meanwhile we note that the estimate (6.3) 
is in the range of values obtained
in two- and three-flavour QCD
from chiral fits of the quark-mass dependence of the pion mass
[\ref{RBCbig}--\ref{tmQCDsigma}]
and from studies of the so-called $\eps$-regime of QCD
[\ref{LangEtAl}--\ref{FukayaIII}].

%% file: figure3
\topinsert
\vbox{
\vskip0.0cm
\centerline{\epsfxsize=8.0cm\epsfbox{plots/nu.eps}}
\vskip0.3cm
\figurecaption{%
Dependence of the renormalized
mode number on $\Mr$ at $\mr\simeq26$ MeV
and $L\simeq2.5$ fm.
The curve shown is based on a representative ensemble of $71$ 
gauge-field configurations and required
the lowest $80$ eigenvalues of $\Dmdag\Dm$ to be calculated for 
each of these fields.
Statistical errors are slightly larger than the jitter of the curve.
}
%\vskip0.3cm
}
\endinsert

%% file: figure4
\topinsert
\vbox{
\vskip0.0cm
\centerline{\epsfxsize=\hsize\epsfbox{plots/nuall.eps}}
\vskip0.3cm
\figurecaption{%
Simulation results for the renormalized mode number at 
fixed $L\simeq2.5$ fm (plot on the left). 
The linear extrapolation to the chiral limit 
(open square)
of $\Sbar$ at $\Mr=95$ MeV is shown on the right.
All errors in these plots are statistical only.
}
%\vskip0.3cm
}
\endinsert

%% file: table1
\topinsert
%Blanke Zahl
\newdimen\digitwidth
\setbox0=\hbox{\rm 0}
\digitwidth=\wd0
\catcode`@=\active
\def@{\kern\digitwidth}
\tablecaption{Simulation results for $\Sbar$ at $\Mr=95$ MeV} 
\vskip1.0ex
$$\vbox{\settabs\+&%
                  xxxxxxx&xx&%     Run
                  xxxxxxxxxxxx&xx&%   m
                  xxxxxxxxxxxxxx&x&\cr%  Sigma
\thicktablerule
\vskip1.0ex
                \+& \hfill Run\hfill
                 && \hfill $\mr$\kern2pt[MeV]\hfill
                 && \hfill $\Sbar^{1/3}$\kern2pt[MeV]\hfill
                 &\cr
\vskip1.0ex
\thintablerule
\vskip1.2ex
  \+& \hfill $E_4$\hfill
  &&  \hfill $45.8(3)(11)$\hfill
  &&  \hfill $310(2)(4)$\hfill
  &\cr
\vskip0.3ex
  \+& \hfill $E_5$\hfill
  &&  \hfill $26.5(2)(6)@$\hfill
  &&  \hfill $295(2)(4)$\hfill
  &\cr
\vskip0.3ex
  \+& \hfill $E_6$\hfill
  &&  \hfill $12.8(2)(3)@$\hfill
  &&  \hfill $286(2)(4)$\hfill
  &\cr
\vskip1.2ex
\thicktablerule
}
$$
%\vskip-0.0ex
\endinsert

%% file: sect7
\section 7. Further uses of spectral observables

Spectral observables like the mode number provide 
interesting probes of low-energy QCD. 
In this section we wish to show that 
the computation of the chiral condensate is only one of the
possible applications of these observables.

\subsection 7.1 Scaling to the continuum limit

Extrapolations to the continuum limit require simulations of 
a series of lattices with decreasing lattice spacings. Since
only the bare coupling and bare quark mass can be prescribed, 
the ratios of the 
spacings of the simulated lattices are not known a priori
and need to be calculated. Evidently, it is very important
to obtain the ratios with small statistical and 
systematic errors.

A set of O($a$)-improved renormalized quantities, 
which may conceivably be used to match the lattices, 
is\kern1.5pt\footnote{$\dagger$}{\footnotefont%
The list of observables given here only serves to illustrate
the general ideas. In particular, 
the combination $\Mr\Sbar$ may be used in place of $\nur/V$.}
\equation{
  \left\{\Mpi,\Mr\Gpir,{\nur\over V}\right\},
  \enum
}
where $\Mpi$ and $\Gpir$ are, respectively, the pion mass and the 
renormalized vacuum-to-pion matrix element of the isospin pseudo-scalar
density. 
All these quantities are renormalization-group
invariants. In particular, the dimensionless combinations
\equation{
  C_1=\Mpi^2\left({V\over\nur}\right)^{1/2},
  \enum
  \next{2.5ex}
  C_2=\left(\Mr\Gpir\right)^4\left({V\over\nur}\right)^3,
  \enum
}
are well-defined and directly accessible functions of $g_0,am_0$ and $aM$.

Since $C_1$ and $C_2$ are roughly linearly rising with $am_0$ and 
$aM$, respectively, it is possible to match the 
mass parameters on any given pair of lattices 
by requiring $C_1$ and $C_2$ to assume the same (sensibly chosen)
values. After that the ratio of the lattice spacings is
obtained through
\equation{
  {a_1\over a_2}=\left(\nu_1n_2\over n_1\nu_2\right)^{1/4},
  \enum
}
where $n_1,n_2$ denote the numbers of points of 
the two lattices and $\nu_1,\nu_2$ the mode numbers
at the matched values of the mass parameters
(we implicitly assumed here that finite-volume
effects can be neglected or that the volumes are the same).

An important technical advantage of this procedure is that all
quantities involved are easily obtained with small errors.
In particular, the statistical precision that can be attained in practice
is not expected to change dramatically as 
the lattice spacing is decreased or if larger lattices
are considered.

\subsection 7.2 Computation of renormalization constants

Density-chain correlation functions like the ones discussed in 
sect.~3 satisfy various chiral Ward identities in the 
continuum limit. We may, for example, start from
the ``twisted spectral sums'' 
\equation{
  \sigma_{k,l}(\mu,\mq)=
  \bigl\langle\Tr\bigl\{\dirac{5}\bigl(\Dmdag\Dm+\mu^2\bigr)^{-k}
  \dirac{5}(\Dmdag\Dm+\mu^2\bigr)^{-l}
  \bigr\}\bigr\rangle,
  \enum
}
which can be represented through
density-chain correlation functions of the form
\equation{
  \sigma_{1,2}(\mu,\mq)=-a^{24}\sum_{x_1,\ldots,x_6}
  \noenum
  \next{2.0ex}
  \hskip7.5em
  \bigl\langle S^{+}_{12}(x_1)P^{-}_{23}(x_2)S^{+}_{34}(x_3)P^{-}_{45}(x_4)
               P^{+}_{56}(x_5)P^{-}_{61}(x_6)\bigr\rangle.
  \enum
}
In the continuum limit and if $k+l\geq3$, 
chiral symmetry (or simply the fact that
$\dirac{5}$ commutes with $\Dmdag\Dm$ in the continuum theory) 
then implies
that the properly renormalized twisted spectral sum 
$\sigma_{k,l,\hbox{\sixrm R}}$ coincides with 
$\sigma_{k+l,\hbox{\sixrm R}}$.

On the lattice one should keep track of the 
$\rmO(a)$ corrections, but following the lines
of ref.~[\ref{Stability}], it is then straightforward 
to show that
\equation{
  {\ZP^2\over\ZS^2}=\left(1+2\bR a\mq\right)
  {\sigma_{k,l}\over\sigma_{k+l}}+\rmO(a^2),
  \enum
  \next{2ex}
  \bR=\bS-\bP+2(\bPP-\bPS),
  \enum
}
where the improvement coefficient $\bR$ is known
to one-loop order of perturbation theory and appears to be 
small (appendix C). 

Equation (7.7) is actually a special case of a more general Ward
identity, where the inverse powers of $\Dmdag\Dm+\mu^2$ in the
definition of the spectral sums are replaced by any
sufficiently rapidly decaying functions of $\Dmdag\Dm$.
In particular,
\equation{
  {\ZP^2\over\ZS^2}=\left(1+2\bR a\mq\right)
  {\langle\Tr\{\dirac{5}\PM\dirac{5}\PM\}\rangle\strut
  \over
  \langle\Tr\{\PM\}\rangle\strut}+\rmO(a^2)
  \enum
}
is an identity recommended for numerical evaluation.

\subsection 7.3 Topological susceptibility

Using parity-odd density chains, the topological susceptibility
$\chi_t$ in QCD can be defined in a manifestly ultraviolet-finite and
therefore universally valid way [\ref{LuscherTop}].
On the lattice there exist different 
definitions of this type, all of which are expected to coincide
in the continuum limit.
In particular, one can make use of twisted-mass density chains 
and it is then possible, as in the case of the chiral Ward identities discussed
in the previous subsection, to pass from density chains to 
spectral projectors.

Proceeding along these lines, the expression
\equation{
  \chi_t=\left(1+2\bR a\mq\right){\ZS^2\over\ZP^2}{1\over V}
  \langle\Tr\{\dirac{5}\PM\}\Tr\{\dirac{5}\PM\}\rangle
  +\rmO(a^2)
  \enum
}
is obtained, which, when combined with eq.~(7.9), leads to the 
formula
\equation{
  \chi_t={\nu\over V}
  {\langle\Tr\{\dirac{5}\PM\}\Tr\{\dirac{5}\PM\}\rangle\strut\over
   \langle\Tr\{\dirac{5}\PM\dirac{5}\PM\}\rangle\strut}
  +\rmO(a^2).
  \enum
}
In principle the mass $\Mr$ can be set to any value in eqs.~(7.9)--(7.11),
but since the size of the lattice effects depends on $\Mr$,
its value should in practice be chosen with some care. 
One evidently requires that $a\Mr\ll1$ and it is certainly wise 
to avoid the threshold region $\Mr\simeq\mr$, where
the lattice effects tend to be kinematically enhanced.
Moreover, a definite prescription that fixes $\Mr$
in physical units should be adopted when
scaling to the continuum limit, as otherwise there is no
guarantee that the calculated renormalized quantities
converge with a rate proportional to $a^2$.

%% file: sect8
\section 8. Concluding remarks

The condensation of the low modes of the Dirac operator
seen in numerical lattice QCD provides 
a most direct piece of theoretical
evidence for the spontaneous breaking of chiral symmetry in QCD.
Explicit violations of chiral symmetry at momenta on the 
order of the inverse lattice spacing
have little influence on the mode condensation, because 
the mode number is a renormalizable quantity 
and therefore coincides with its continuum limit
up to terms that vanish proportionally to a power of 
the lattice spacing.

The dynamical mechanisms that cause the modes to condense
are presently not known. It is quite clear, however, that 
the spontaneous breaking of chiral symmetry is not 
a many-quark collective effect. 
The mode condensation actually appears to be largely insensitive to
the sea-quark mass and it seems to persist even when 
passing to the quenched theory.
There is thus no relevant back-reaction of the sea quarks
and theoretical studies of the behaviour of a single quark
in presence of representative gauge fields may therefore
allow the breaking of chiral symmetry to be explained.

While the computation of the chiral condensate is an obvious application of 
the spectral projector technique introduced in this paper,
there are other applications as well and the technique is also not 
limited to a particular lattice formulation of QCD. Moreover, it may
be useful for studies of the theory at non-zero temperature 
and of QCD-like theories, 
where chiral symmetry may or may not 
be spontaneously~broken.

\vskip1.0ex

We wish to thank Stefan Sint and Peter Weisz for 
correspondence on the O($a$)-improvement of twisted-mass QCD and
Filippo Palombi for his help in producing the eigenvalue data 
on which fig.~3 is based.
The gauge-field configurations used for the numerical studies 
were provided by the CLS community [\ref{CLS}].
All computations were performed on PC clusters at CERN and CILEA.
We are grateful to these institutions for providing the required resources
and their technical staff for assistance.

%% file: appa
\appendix A. SU($\bf3|1$) chiral perturbation theory

As explained in sect.~4, the chiral expansion
of the spectral density (and thus of the mode number) 
is obtained by calculating the valence-quark condensate $\Sval(\mval)$ in
partially quenched chiral perturbation theory
[\ref{BernardGolterman},\ref{SharpeShoresh}].
We here provide some details of this computation,
assuming the reader is familiar with chiral
perturbation theory and partial quenching.

Following a suggestion of Sharpe and Shoresh [\ref{SharpeShoresh}], 
we do not include a flavour-singlet field in the effective chiral
theory. In the sea-quark sector, the chiral expansions generated
by the ${\rm SU}(3|1)$ chiral lagrangian then literally coincide with 
those obtained in the standard SU(2) theory.
In particular, the low-energy constants (such as $F$ and $\Sigma$) 
which already occur in the latter are the same.

\subsection A.1 Group generators 

The complex Lie superalgebra of SU($3|1$) consists of all $4\times 4$
supermatrices $X_{\alpha\beta}$ with vanishing supertrace (see
ref.~[\ref{Cornwell}], for example). We assume the indices 
$\alpha,\beta$ of these
matrices to be such that $\alpha=1,2$ corresponds to the sea quarks,
$\alpha=3$ to the valence quark and $\alpha=4$ to the ghost (or
pseudo-fermion) quark associated to the valence quark.

Our conventions for the generators 
$T^a$, $a=1,\ldots,15$, of the algebra are
\equation{
  T^a=(T^a)^{\dagger},
  \qquad
  \Str\{T^a\}=0, 
  \qquad 
  \Str\,\{T^aT^b\}=\frac{1}{2}g^{ab},
  \enum
}
where the non-zero elements of the
matrix $g^{ab}$ are given by
\equation{
  g=\pmatrix{1\cr
               &\ddots\cr
               &      &1\cr
               &      & &-\tau^2\cr
               &      & &       &\ddots\cr
               &      & &       &      &-\tau^2\cr
               &      & &       &      &       &-1\cr}
  \matrix{\left.\vphantom{\matrix{1\cr
                                   &\ddots\cr
                                   &      &1\cr}}\right\}&
                                   \kern-1.5ex1-8\hfill\cr
          \left.\vphantom{\matrix{-\tau^2\cr
                                         &\ddots\cr
                                         &      &-\tau^2\cr}}\right\}&
                                   \kern-1.5ex9-14\hfill\cr
          \left.\vphantom{\matrix{1\cr}}\right\}&
                                   \kern-1.5ex15\hfill\cr
         }
  \enum
}
More specifically, $T^1,\ldots,T^8$ are assumed to be
generators of the SU(3) subgroup acting on the sea and valence quarks,
while $T^9,\ldots,T^{14}$ mix the ghost with the other quarks
and $T^{15}$ is a diagonal matrix with a non-zero ghost-quark
component.

In the following 
the Einstein summation convention is adopted for 
SU($3|1$) group indices and for Lorentz indices.
It is also helpful to introduce the tensors
\equation{
  h^{ab}=(g^{a8}-g^{a15})(g^{b8}-g^{b15}), 
  \qquad
  k^{ab}=(g^{a8}+g^{a15})(g^{b8}+g^{b15}), 
  \enum
}
which satisfy $k^{ac}h^{cb}=0$.

\subsection A.2 Chiral effective lagrangian

The ${\rm SU}(3|1)$ 
chiral effective theory
is a non-linear $\sigma$-model in which the basic field $U(x)$ 
takes values in ${\rm SU}(3|1)$. As usual the lagrangian
\equation{
  \Leff=\Leff^{(2)}+\Leff^{(4)}+\ldots
  \enum
}
is given as a series of terms of increasing dimension.
The leading-order term,
\equation{
  {\cal L}^{(2)}=
  -\frac{1}{4}F^2\kern1pt\Str\{J_{\mu}J_{\mu}\}
  -\frac{1}{2}BF^2\kern1pt\Str\{MU^{\dagger}+M^{\dagger}U\},
  \qquad 
  J_{\mu}=U^{\dagger}\partial_{\mu}U,
  \enum 
}
involves the quark mass matrix $M$, the pion decay constant 
in the chiral limit, $F$,
and the parameter $B$, which is related to the quark condensate
through $\Sigma=BF^2$.
The mass matrix is taken to be diagonal,
\equation{
   M={\rm diag}\kern1pt\{m,m,\mval,\mvalt\},  
   \enum
}
where $m$, $\mval$ and $\mvalt$ are, respectively, the masses of the sea quarks, 
the valence quark and the ghost quark. In order to properly 
quench the valence quark, $\mvalt$ will later be set to $\mval$.

At next-to-leading order, the effective lagrangian reads
\equation{
  {\cal L}^{(4)}=  
  -L_0\kern1pt
   \Str\{J_{\mu} J_{\nu} J_{\mu}J_{\nu}\}
  -(L_1-\frac{1}{2}L_0)\kern1pt
   \Str\{J_{\mu}J_{\mu}\}\kern1pt\Str\{J_{\nu}J_{\nu}\}
  \noenum 
  \next{2.5ex}
   {\phantom{{\cal L}^{(4)}=}}
   -(L_2-L_0)\kern1pt
    \Str\{J_{\mu}J_{\nu}\}\kern1pt\Str\{J_{\mu}J_{\nu}\} 
   -(L_3+2L_0)\kern1pt
    \Str\{J_{\mu}J_{\mu}J_{\nu}J_{\nu}\}
  \noenum
  \next{2.5ex}
  {\phantom{{\cal L}^{(4)}=}}
  -2BL_4\kern1pt 
   \Str\{J_{\mu}J_{\mu}\}\kern1pt\Str\{MU^{\dagger}+M^{\dagger}U\}
  -2BL_5\kern1pt 
   \Str\{J_{\mu}J_{\mu}(U^{\dagger}M+M^{\dagger}U)\}
  \noenum
  \next{2.5ex} 
  {\phantom{{\cal L}^{(4)}=}}
  -4B^2L_6\kern1pt
   \Str\{U^{\dagger}M+M^{\dagger}U\}\kern1pt\Str\{U^{\dagger}M+M^{\dagger}U\}
  \noenum  
  \next{2.5ex}
  {\phantom{{\cal L}^{(4)}=}}
  -4B^2L_7\kern1pt 
   \Str\{M^{\dagger}U-MU^{\dagger}\}\kern1pt\Str\{M^{\dagger}U-MU^{\dagger}\}
  \noenum  
  \next{2.5ex} 
  {\phantom{{\cal L}^{(4)}=}}
  -4B^2L_8\kern1pt
   \Str\{MU^{\dagger}MU^{\dagger}+M^{\dagger}UM^{\dagger}U\}
  -4B^2H_2\kern1pt 
   \Str\{M^{\dagger}M\}.
  \enum
}
The additional low-energy constants at this order are thus 
$L_0,\ldots,L_8$ and $H_2$. Note that in these expressions 
we have omitted all terms that do not contribute to the 
valence-quark condensate (such as those related
to current correlation functions, for example)
[\ref{GasserLeutwyler}].

\subsection A.3 Perturbation expansion

The chiral expansion of 
\equation{
  \Sval(\mval)=-\left\langle\sigma_{\rm val}\right\rangle_{\mvalt=\mval},
  \enum
  \next{2.5ex}
  \sigma_{\rm val}={\partial\Leff\over\partial\mval}
  =-\Sigma\kern1pt\Re U_{33}+\ldots,
  \enum
}
is obtained by substituting
\equation{
  U=\exp\left\{2i\phi/F\right\},   
  \qquad\phi=\phi^aT^a, 
  \enum
}
in the functional integral and expanding all entries in powers of 
$\phi$. Since the expectation value in eq.~(A.8) is to be computed
at $\mvalt=\mval$, one needs 
to work out the Feynman rules only for this case. 

To second order in $\phi$, the leading-order lagrangian reads
\equation{
  {\cal L}^{(2)}=
  \frac{1}{2}g^{ab}\{\partial_\mu\phi^a\partial_\mu\phi^b+
  M_a^2\phi^a\phi^b\}
  +\frac{1}{6}(\Mss^2-\Mvv^2)k^{ab}\phi^a\phi^b, 
  \enum
}
where $\Mss^2=2Bm$, $\Mvv^2=2B\mval$ and
\equation{
  M^2_a
  =\cases{\Mss^2                     & if $a=1,2,3$,\cr
          \noalign{\vskip1.0ex}
          \frac{1}{2}(\Mss^2+\Mvv^2) & if $a=4,\dots,7,9,\dots,12$,\cr
          \noalign{\vskip1.0ex}
          \Mvv^2                     & if $a=8,13,14,15$.\cr}  
  \enum
}
The propagator of the meson field is thus given by
\equation{
  \langle\phi^a(x)\phi^b(0)\rangle=g^{ab}\Prop_1(x,M_a^2)+
  \frac{1}{3}(\Mvv^2-\Mss^2) 
  h^{ab}\Prop_2(x,M_a^2), 
  \enum
  \next{2.5ex}
  \Prop_n(x,M^2)= 
  \int{\rmd^4 p\over(2\pi)^4}\,{\rme^{ipx}\over(p^2+M^2)^n}.
  \enum
}
All other Feynman rules can be derived straightforwardly 
from the lagrangian and the field $\sigma_{\rm val}$.

Following common practice, we use dimensional regularization for 
the loop integrals and a modified minimal subtraction scheme
for the bare couplings in the lagrangian $\Leff^{(4)}$.  
In particular, in $4-2\eps$ dimensions we substitute
\equation{
  L_6={3\mubar^{-2\eps}\over(32\pi)^2}\left\{-{1\over\eps}
  +\gamma-\ln4\pi-1+\lbar_6\right\}
  \enum
}
for the coupling $L_6$,
where $\gamma=0.577\ldots$ denotes Euler's constant, $\lbar_6$
the renormalized coupling and $\mubar$ the renormalization scale.

\subsection A.4 Finite-volume correction

The chiral expansion
\equation{
  \Sval(\mval)-\left.\Sval(\mval)\right|_{V=\infty}=
  \noenum
  \next{2.5ex} 
  \hskip0.5em{\Sigma\over2F^2} 
  \left\{g_1\bigl(\Mvv^2\bigr)
         -4g_1\bigl(\frac{1}{2}\Mss^2+\frac{1}{2}\Mvv^2\bigr) 
         +\left(\Mss^2-\Mvv^2\right)g_2\bigl(\Mvv^2\bigr)\right\}
  +\ldots
  \enum
}
starts at next-to-leading order and involves the momentum sums
\equation{
  g_n(M^2)={1\over V}\sum_p{1\over(p^2+M^2)^n}-\Prop_n(0,M^2).
  \enum
}
These are easily calculated numerically when written  
in the form of rapidly converging series of
Bessel functions [\ref{GasserLeutwylerFV}].

%% file: appb
\appendix B. Estimation of the approximation error $\bf\Delta$

The computational strategy outlined in sect.~5 assumes that 
the parameters $n$, $\eps$, $M$ and $\Mstar$ are such that the
approximation error $\Delta$ [eq.~(5.8)] can be safely neglected. 
In this appendix, we now show how this condition can be met
in practice.

\subsection B.1 Spectral integral

Our starting point is the spectral integral representation
\equation{
  \Delta=
  \int_{0}^{\infty}\rmd\omega\,
  \left\{\theta(M-\omega)-h(x_{\omega})^4\right\}\nup(\omega,\mq)
  \enum
}
in which
\equation{
  \nup(\omega,\mq)={\partial\over\partial\omega}\nu(\omega,\mq),
  \qquad
  x_{\omega}=1-{2\Mstar^2\over\omega^2+\Mstar^2}.
  \enum
}
Note that $\nup(\omega,\mq)$ coincides with the average 
spectral density of the square root of $\Dmdag\Dm$ up to a 
factor $V$. For illustration, 
the two functions in the curly bracket 
are plotted in fig.~5 for a typical choice of the parameters.

In the following,
we distinguish three ranges of $\omega$, separated by the limits
\equation{
  \omega_{\pm}=
  \Mstar\left({1\pm\sqrt{\eps}\over1\mp\sqrt{\eps}}\right)^{1/2}
  \enum
}
of the transition region around $\omega=\Mstar$ (see fig.~5).
The parts of the spectral integral corresponding to the 
integration ranges $[0,\omm]$, $[\omm,\omp]$ and $[\omp,\infty]$
are denoted by $\delm$, $\delt$ and $\delp$, respectively.

\input figure5

\subsection B.2 Bounds on $\delp$ and $\delm$

Noting
\equation{
  x_{\omega}=\pm\sqrt{\eps}\quad\hbox{at}\quad\omega=\ompm
  \enum
}
and recalling the approximation property (5.5) of the minmax
polynomial $P(y)$, the function in the curly bracket
in eq.~(B.1) is easily bounded in the case of the integrals $\delpm$.
Since the total number of eigenmodes of $\Dmdag\Dm$ is 
$12 V/a^4$ and since there are at most $\nu(M,\mq)$
eigenmodes with eigenvalues $\omega^2\leq\omm^2$,
it is then straightforward to establish the bounds
\equation{
  |\delp|\leq{3\over4}\kern0.5pt{V\delta^4\over a^4},
  \enum
  \next{2.5ex}
  |\delm|\leq\nu(M,\mq)\left\{2\delta+\rmO(\delta^2)\right\}.
  \enum
}
These parts of the total error $\Delta$ are thus controlled by the 
precision $\delta$ of the polynomial approximation
to the step function.

In ref.~[\ref{NumMethods}] it was shown that $\delta$ is an
exponentially decreasing function of $n\sqrt{\eps}$.
The precision can therefore be set to the desired level
by adjusting the degree $n$ of the minmax polynomial. 
If $\eps=0.01$, for example, and if a lattice of size
$128\times64^3$ is considered, a sensible choice is $n=32$ 
and eqs.~(B.5),(B.6) then imply that
$|\delp|\leq10^{-6}$ and $|\delm|\leq10^{-3}\times\nu(M,\mq)$.

\subsection B.3 Estimation of $\delt$ and the relation of $M$ to $\Mstar$

The remaining error component, $\delt$, is more difficult to estimate
than $\delp$ and $\delm$.
An important point to note is that the density
$\nup(\omega,\mq)$ tends to be practically constant 
in the transition region (cf.~sect.~6).
Most of the error can therefore be cancelled
by choosing the relation between $M$ and $\Mstar$ 
to be such that
$\delt$ vanishes for a constant density.
This condition amounts to setting
\equation{
  {M\over\Mstar}=
  \left({1-\sqrt{\eps}\over1+\sqrt{\eps}}\right)^{1/2}
  +\int_{-\sqrt{\eps}}^{\sqrt{\eps}}
  \rmd x\,{1+x\over(1-x^2)^{3/2}}\,h(x)^4
  \enum
}
and the residual value of the error,
\equation{
  \delt=
  \int_{\omm}^{\omp}\rmd\omega\,
  \left\{\theta(M-\omega)-h(x_{\omega})^4\right\}
  \left\{\nup(\omega,\mq)-\nup(M,\mq)\right\},
  \enum
}
is then of order $\eps$. 

An estimation of $\delt$ however requires 
some information on the $\omega$-dependence
of the density $\nup(\omega,\mq)$ in the transition region.
For a determination of the expected order of magnitude of 
$\delt$, chiral perturbation theory may be used at this point
and a rough bound on the slope of $\nup(\omega,\mq)$
(and thus on $\delt$) can also be obtained a posteriori 
through a fit of simulation results for the mode number.
Whether $\delt$ is in fact negligible with respect to the statistical
errors can ultimatly always
be checked by varying $\eps$ at fixed $\delta$.

%% file: figure5
\topinsert
\vbox{
\vskip0.0cm
\centerline{\epsfxsize=8.5cm\epsfbox{plots/stepfct.eps}}
\vskip0.3cm
\figurecaption{%
Approximate spectral step function $h(x_{\omega})^4$ for
$n=32$, $\eps=0.01$ and $\Mstar=94$ MeV.
The exact step function $\theta(M-\omega)$ 
is also shown (grey line), where $M$ and $\Mstar$ 
are related through eq.~(B.7). 
}
%\vskip0.3cm
}
\endinsert

%% file: appc
\appendix C. Lattice parameters and simulation results

\vskip-2.5ex

\subsection C.1 Lattice parameters

The numerical studies reported in sect.~6
are based on representative ensembles of gauge-field configurations
for the two-flavour O($a$)-improved Wilson theory (cf.~subsect.~2.2). 
The ensembles
were generated by the authors of ref.~[\ref{CERNTOV}] and
were made available to us through the CLS community effort [\ref{CLS}].

\input table2

In these simulations, the coupling $\beta=6/g_0^2$
was set to $5.3$ in all cases and the sea-quark hopping parameter
$\kappa=(8+2m_0)^{-1}$ to the values quoted in table~2.
The lattice sizes and the numbers $N_{\rm cfg}$ of configurations
are also given in the table.
The spacing of the two lattices considered
was determined to be $0.0784(10)$ fm [\ref{CERNTOV}]
and their spatial sizes are thus 
$L=1.88(2)$ fm and $L=2.51(3)$ fm, respectively.

\subsection C.2 Computation of the mode number

The mode numbers listed in table~2 were computed
stochastically following the lines of sect.~5.
We used the same minmax polynomial of degree $n=32$
in all cases, with $\eps$ set to $0.01$, and the number
$N$ of pseudo-fermion fields was taken to be $1$. 
With these choices, $\delta=4.4\times10^{-4}$,
the integral (B.7) evaluates to $M/\Mstar=0.96334$
and the approximation error $\Delta$ [eq.~(5.8)] is
estimated to be neglible in our computations.

The statistical errors quoted in table~2 are in the range
from $1$ to $2$ percent. 
They are practically given by 
the last term in eq.~(5.4), which explains why,
with half the statistics on the larger lattices, approximately
the same relative accuracy is obtained on all lattices.
Moreover, the errors could be further reduced, by 
a factor $2$ at least,
by increasing the number $N$ of pseudo-fermion fields.

\subsection C.3 $\rmO(a)$-improvement and renormalization at $\beta=5.3$

The coefficients of the O($a$) counterterms in the
quark action [\ref{SW}] and the improved axial current [\ref{OaImp}]
were set to the non-perturbatively determined values
$\csw=1.90952$ [\ref{NPimp}]
and $\cA=-0.0506$ [\ref{DellaMorteCA}],
respectively. We computed the renormalized quark mass via the 
PCAC relation, using the improved axial current, 
but neglected the O($a\mq$) corrections in 
eq.~(2.8) since $\bA-\bP=-0.00104(6)\times g_0^2+\rmO(g_0^4)$
is very small [\ref{OaOneLoopII}]. 

Although a different improvement scheme was adopted in 
ref.~[\ref{tmQCDimp}], it is possible to deduce
the one-loop formulae
\equation{
  \bmu=-\frac{1}{2}-0.111(4)\times g_0^2+\rmO(g_0^4),
  \enum
  \next{2.5ex}
  \bR=-0.031(8)\times g_0^2+\rmO(g_0^4),
  \enum
}
from the results published there and in 
refs.~[\ref{OaOneLoopI},\ref{OaOneLoopII}]. So far 
$\bmu$ is only known in perturbation theory and we thus
used the one-loop estimate $\bmu=-0.626$ in eq.~(3.11).
Noting $a\mc=-0.33560(5)$, the subtracted bare mass $a\mq$
is smaller than $0.01$ at all values of $\kappa$ considered.
The calculated O($a\mq$) corrections to $\Mr$
are therefore at most $0.6\%$ and the corrections to
the ratio (7.9) will normally be negligible.

The renormalization factors 
$\ZA=0.778(10)$ [\ref{DellaMorteZA}] and
$\ZP=0.543(8)$ [\ref{DellaMorteZPI},\ref{DellaMorteZPII}]
needed to pass from the bare masses $m$ and $M$ to
the renormalized masses $\mr$ and $\Mr$
in the $\MSbar$ scheme at $2$ GeV
have been computed non-perturbatively.
As can be seen from table~1, the renormalized sea-quark mass ranges
from about $13$ to $46$ MeV on the lattices considered.
The values of $aM$ in table~2 have, incidentally, been chosen
such that $\Lr=(\Mr^2-\mr^2)^{1/2}$ approximately assumes the values
$70,85,100$ and $115$ MeV
at all sea-quark masses.

%% file: table2
\topinsert
%Blanke Zahl
\newdimen\digitwidth
\setbox0=\hbox{\rm 0}
\digitwidth=\wd0
\catcode`@=\active
\def@{\kern\digitwidth}
\tablecaption{Simulation results for the mode number} 
\vskip0.0ex
$$\vbox{\settabs\+&%
                  xxxxxxx&xx&%     Run
                  xxxxxxxxx&xx&%   Lattice
                  xxxxxxxxx&xx&%   Kappa
                  xxxxxxxxx&xx&%   Ncfg		   
                  xxxxxxxxx&xx&%   M
                  xxxxxxxxxx&x&\cr%  nu
\thicktablerule
\vskip1.0ex
                \+& \hfill Run\hfill
                 && \hfill Lattice\hfill
                 && \hfill $\kappa$\hfill
                 && \hfill $N_{\rm cfg}$\hfill
                 && \hfill $aM$\hfill
                 && \hfill $\nu(M,\mq)$\hfill
                 &\cr
\vskip1.0ex
\thintablerule
\vskip1.2ex
  \+& \hfill $D_3$\hfill
  &&  \hfill $48\times24^3$\hfill
  &&  \hfill $0.13610$\hfill
  &&  \hfill $160$\hfill
  &&  \hfill $0.02674$\hfill 
  &&  \hfill $28.6(4)@$\hfill
  &\cr
\vskip0.3ex
  \+& \hfill \hfill
  &&  \hfill $$\hfill
  &&  \hfill \hfill
  &&  \hfill $$\hfill
  &&  \hfill $0.02377$\hfill 
  &&  \hfill $24.0(4)@$\hfill
  &\cr
\vskip0.3ex
  \+& \hfill \hfill
  &&  \hfill $$\hfill
  &&  \hfill \hfill
  &&  \hfill $$\hfill
  &&  \hfill $0.02087$\hfill 
  &&  \hfill $19.5(4)@$\hfill
  &\cr
\vskip0.3ex
  \+& \hfill \hfill
  &&  \hfill $$\hfill
  &&  \hfill \hfill
  &&  \hfill $$\hfill
  &&  \hfill $0.01808$\hfill 
  &&  \hfill $15.2(3)@$\hfill
  &\cr
\vskip1.0ex
\thintablerule
\vskip1.2ex
  \+& \hfill $D_5$\hfill
  &&  \hfill $48\times24^3$\hfill
  &&  \hfill $0.13625$\hfill
  &&  \hfill $160$\hfill
  &&  \hfill $0.02549$\hfill 
  &&  \hfill $26.8(5)@$\hfill
  &\cr
\vskip0.3ex
  \+& \hfill \hfill
  &&  \hfill $$\hfill
  &&  \hfill \hfill
  &&  \hfill $$\hfill
  &&  \hfill $0.02234$\hfill 
  &&  \hfill $22.6(4)@$\hfill
  &\cr
\vskip0.3ex
  \+& \hfill \hfill
  &&  \hfill $$\hfill
  &&  \hfill \hfill
  &&  \hfill $$\hfill
  &&  \hfill $0.01923$\hfill 
  &&  \hfill $18.5(4)@$\hfill
  &\cr
\vskip0.3ex
  \+& \hfill \hfill
  &&  \hfill $$\hfill
  &&  \hfill \hfill
  &&  \hfill $$\hfill
  &&  \hfill $0.01616$\hfill 
  &&  \hfill $14.4(3)@$\hfill
  &\cr
\vskip1.0ex
\thintablerule
\vskip1.2ex
  \+& \hfill $E_4$\hfill
  &&  \hfill $64\times32^3$\hfill
  &&  \hfill $0.13610$\hfill
  &&  \hfill $80$\hfill
  &&  \hfill $0.02674$\hfill 
  &&  \hfill $93.9(12)$\hfill
  &\cr
\vskip0.3ex
  \+& \hfill \hfill
  &&  \hfill $$\hfill
  &&  \hfill \hfill
  &&  \hfill $$\hfill
  &&  \hfill $0.02377$\hfill 
  &&  \hfill $78.2(12)$\hfill
  &\cr
\vskip0.3ex
  \+& \hfill \hfill
  &&  \hfill $$\hfill
  &&  \hfill \hfill
  &&  \hfill $$\hfill
  &&  \hfill $0.02087$\hfill 
  &&  \hfill $63.4(10)$\hfill
  &\cr
\vskip0.3ex
  \+& \hfill \hfill
  &&  \hfill $$\hfill
  &&  \hfill \hfill
  &&  \hfill $$\hfill
  &&  \hfill $0.01808$\hfill 
  &&  \hfill $48.9(9)@$\hfill
  &\cr
\vskip1.0ex
\thintablerule
\vskip1.2ex
  \+& \hfill $E_5$\hfill
  &&  \hfill $64\times32^3$\hfill
  &&  \hfill $0.13625$\hfill
  &&  \hfill $80$\hfill
  &&  \hfill $0.02549$\hfill 
  &&  \hfill $83.2(10)$\hfill
  &\cr
\vskip0.3ex
  \+& \hfill \hfill
  &&  \hfill $$\hfill
  &&  \hfill \hfill
  &&  \hfill $$\hfill
  &&  \hfill $0.02234$\hfill 
  &&  \hfill $69.6(9)@$\hfill
  &\cr
\vskip0.3ex
  \+& \hfill \hfill
  &&  \hfill $$\hfill
  &&  \hfill \hfill
  &&  \hfill $$\hfill
  &&  \hfill $0.01923$\hfill 
  &&  \hfill $56.4(8)@$\hfill
  &\cr
\vskip0.3ex
  \+& \hfill \hfill
  &&  \hfill $$\hfill
  &&  \hfill \hfill
  &&  \hfill $$\hfill
  &&  \hfill $0.01616$\hfill 
  &&  \hfill $44.6(7)@$\hfill
  &\cr
\vskip1.0ex
\thintablerule
\vskip1.2ex
  \+& \hfill $E_6$\hfill
  &&  \hfill $64\times32^3$\hfill
  &&  \hfill $0.13635$\hfill
  &&  \hfill $80$\hfill
  &&  \hfill $0.02499$\hfill 
  &&  \hfill $78.7(11)$\hfill
  &\cr
\vskip0.3ex
  \+& \hfill \hfill
  &&  \hfill $$\hfill
  &&  \hfill \hfill
  &&  \hfill $$\hfill
  &&  \hfill $0.02177$\hfill 
  &&  \hfill $66.6(11)$\hfill
  &\cr
\vskip0.3ex
  \+& \hfill \hfill
  &&  \hfill $$\hfill
  &&  \hfill \hfill
  &&  \hfill $$\hfill
  &&  \hfill $0.01856$\hfill 
  &&  \hfill $54.9(9)@$\hfill
  &\cr
\vskip0.3ex
  \+& \hfill \hfill
  &&  \hfill $$\hfill
  &&  \hfill \hfill
  &&  \hfill $$\hfill
  &&  \hfill $0.01537$\hfill 
  &&  \hfill $43.8(8)@$\hfill
  &\cr
\vskip1.2ex
\thicktablerule
}
$$
\vskip-0.0ex
\endinsert

%% file: biblio
\beginbibliography

% Banks-Casher relation

\bibitem{BanksCasher}
T. Banks, A. Casher,
Nucl. Phys. B169 (1980) 103

% Wilson formulation of lattice QCD

\bibitem{Wilson}
K. G. Wilson, Phys. Rev. D10 (1974) 2445

% O(a) improved lattice QCD

\bibitem{SW}
B. Sheikholeslami, R. Wohlert,
Nucl. Phys. B259 (1985) 572

\bibitem{OaImp}
M. L\"uscher, S. Sint, R. Sommer, P. Weisz,
Nucl. Phys. B478 (1996) 365

% Stability paper

\bibitem{Stability}
L. Del Debbio, L. Giusti, M. L\"uscher, R. Petronzio, N. Tantalo,
JHEP 0602 (2006) 011

% Non-perturbative determination of csw

\bibitem{NPimp}
K. Jansen, R. Sommer (ALPHA collab.),
Nucl. Phys. B530 (1998) 185
[E: {\it ibid.} B643 (2002) 517]

% Twisted-mass QCD

\bibitem{tmQCD}
R. Frezzotti, P. A. Grassi, S. Sint, P. Weisz,
JHEP 0108 (2001) 058

\bibitem{tmQCDimp}
R. Frezzotti, S. Sint, P. Weisz,
JHEP 0107 (2001) 048

% Density chains and the topological charge

\bibitem{GiustiRossiTesta}
L. Giusti, G. C. Rossi, M. Testa,
Phys. Lett. B587 (2004) 157

\bibitem{LuscherTop}
M. L\"uscher,
Phys. Lett. B593 (2004) 296

% Spectral density to one-loop order of ChPT

\bibitem{SmilgaStern}
A. Smilga, J. Stern,
Phys. Lett. B318 (1993) 531

\bibitem{OsbornEtAl}
J. C. Osborn, D. Toublan, J. J. M. Verbaarschot,
Nucl. Phys. B540 (1999) 317

\bibitem{DamgaardEtAl}
P. H. Damgaard, J. C. Osborn, D. Toublan, J. J. M. Verbaarschot,
Nucl. Phys. B547 (1999) 305

% Partially quenched ChPT

\bibitem{BernardGolterman}
C. W. Bernard, M. F. L. Golterman, Phys. Rev. D49 (1994) 486

\bibitem{SharpeShoresh}
S. R. Sharpe, N. Shoresh, Phys. Rev. D64 (2001) 114510

% SU(2) ChPT

\bibitem{GasserLeutwyler}
J. Gasser, H. Leutwyler,
Ann. Phys. 158 (1984) 142

% ChPT in finite volume

\bibitem{GasserLeutwylerFV}
J. Gasser, H. Leutwyler,
Phys. Lett. B184 (1987) 83; {\it ibid.} B188 (1987) 477;
Nucl. Phys. B307 (1988) 763

% Eigenvalue fluctuations

\bibitem{LuscherPalombi}
M. L\"uscher, F. Palombi,
PoS (LATTICE 2008) 049

% Minmax polynomial

\bibitem{NumMethods}
L. Giusti, C. Hoelbling, M. L\"uscher, H. Wittig,
Comput. Phys. Commun. 153 (2003) 31

% Clenshaw recursion

\bibitem{Recipes}
W. H. Press, S. A. Teukolsky, W. T. Vetterling, B. P. Flannery,
Numerical recipes in FORTRAN, 2nd ed. (Cambridge University Press,
Cambridge, 1992)

% Discretisation effects on the spectral density

\bibitem{Sharpe}
S. R. Sharpe,
Phys. Rev. D74 (2006) 014512

% Determinations of Sigma from the GMOR relation

\bibitem{RBCbig}
C. Allton et al. (RBC and UKQCD collab.),
arXiv:0804.0473v1 [hep-lat]

\bibitem{NoakiEtAl}
J. Noaki et al. (JLQCD and TWQCD collab.),
Phys. Rev. Lett. 101 (2008) 202004

\bibitem{tmQCDsigma}
P.~Dimopoulos et al. (ETM collab.),
PoS (LATTICE 2008) 103

% Epsilon-regime estimates of Sigma

\bibitem{LangEtAl}
C. B. Lang, P. Majumdar, W. Ortner,
Phys. Lett. B649 (2007) 225

\bibitem{FukayaI}
H. Fukaya et al. (JLQCD collab.),
Phys. Rev. Lett. 98 (2007) 172001

\bibitem{FukayaII}
H. Fukaya et al. (JLQCD and TWQCD collab.),
Phys. Rev. D76 (2007) 054503

\bibitem{HasenfratzSigma}
P. Hasenfratz et al.,
arXiv:0707.0071v2 [hep-lat]

\bibitem{FukayaIII}
H. Fukaya et al. (JLQCD collab.),
Phys. Rev. D77 (2008) 074503

% References quoted in the appendices

% Superalgebras

\bibitem{Cornwell}
J. F. Cornwell,
Group Theory in Physics, Vol. 3
(Academic Press, London, 1989).

% CLS runs

\bibitem{CERNTOV}
L. Del Debbio, L. Giusti, M. L\"uscher, R. Petronzio, N. Tantalo,
JHEP 0702 (2007) 056 and 082

\bibitem{CLS}
{\tt https://twiki.cern.ch/twiki/bin/view/CLS/WebHome}

% Improvement coefficient cA

\bibitem{DellaMorteCA}
M. Della Morte, R. Hoffmann, R. Sommer,
JHEP 0503 (2005) 029

% b-coefficients

\bibitem{OaOneLoopI}
M. L\"uscher, P. Weisz,
Nucl. Phys. B479 (1996) 429

\bibitem{OaOneLoopII}
S. Sint, P. Weisz,
Nucl. Phys. B502 (1997) 251;
Nucl. Phys. (Proc.Suppl.) 63 (1998) 856

% Renormalization constant ZA

\bibitem{DellaMorteZA}
M. Della Morte, R. Sommer, S. Takeda,
arXiv:0807.1120v2 [hep-lat]

% Renormalization constant ZP

\bibitem{DellaMorteZPI}
M. Della Morte et al. (ALPHA collab.),
Nucl. Phys. B 729 (2005) 117

\bibitem{DellaMorteZPII}
M. Della Morte et al. (ALPHA collab.),
JHEP 0807 (2008) 037

\endbibliography